\documentclass[aps,prl,groupedaddress,showpacs,showkeys,twocolumn,amsmath,amsfonts]{revtex4-1}
\usepackage{graphicx}

\begin{document}


\title{Microcanonical entropy for classical systems}


\author{Roberto Franzosi}
\email[]{roberto.franzosi@ino.it}
\affiliation{QSTAR \& CNR - Istituto Nazionale di Ottica, Largo Enrico Fermi 2, I-50125 Firenze, Italy}


\date{\today}

\begin{abstract}
The entropy definition in the microcanonical ensemble is revisited. 
We propose a novel definition for the microcanonical
entropy that resolve the debate
on the correct definition of the microcanonical entropy. In particular we
show that this entropy definition fixes the problem inherent the exact
extensivity of the caloric equation.
Furthermore, this entropy reproduces results which are in agreement
with the ones predicted with standard Boltzmann entropy when applied to
macroscopic systems. 
On the contrary, the predictions obtained with the standard Boltzmann
entropy and with the entropy we propose, are different for
small system sizes. Thus, we conclude that the Boltzmann entropy
provides a correct description for macroscopic systems whereas
extremely small systems should be better described with the entropy
that we propose here.

\end{abstract}


\maketitle

In many of statistical physics applications the entropy notion enters as
a  basic concept suitable for characterize the behaviour of macroscopic
systems \cite{entropy_economic,tsallis2004,entropy_ecology,Franzosi_EPL15,
Franzosi_PRE16,Felice_PhysA_2018}.
In the present manuscript, we address the problem of the correct definition
of the microcanonical entropy for classical systems. In fact, the latter
concern has recently become a matter of a debate where it has been discussed
which one between the Boltzmann and the Gibbs definition provides the correct
entropy. 

A mechanically and adiabatically isolated system, at the equilibrium and
composed of a macroscopic number of interacting particles is statistically
described with the microcanonical ensemble.
In this statistic description the relevant thermodynamical quantities are derived
from the entropy $S$ through suitable thermodynamic relations.
Now, there are -at least- two accepted definitions for the microcanonical entropy,
the ones commonly referred to as Boltzmann entropy and Gibbs entropy.
The former is proportional to the logarithm
of the density of microstates at a given ``energy shell'', whereas the latter 
is proportional to the logarithm of the number of microstates up to a given
energy.
The debate as to which of these definitions of entropy is the correct one dates back
to many years ago
\cite{Hertz10,Einstein11,Schluter48,Jaynes,Munster_1987,Pearson85,Berdichevsky91,Adib04,
Lavis2005245,Campisi05}. 

Very recently \cite{Dunkel2013,Hilbert_PRE_2014,Hanggi_15}, it has been argued that
the Gibbs entropy yields a consistent thermodynamics, and they have been discussed some
consistency issues that, the microcanonical statistical mechanics founded on the
Boltzmann entropy, would unveil 
\cite{Dunkel2013,Hilbert_PRE_2014,Sokolov_2014,DunkelHilbertRep1,DunkelHilbertRep2,
Campisi_2015,Campisi_2016}. These and other related arguments 
\cite{Romero-Rochin,Treumann_2014,Treumann_2014a} have
been contended \cite{Vilar_2014, Frenkel_2015, Schneider_2014,Wang_2015, Cerino_2015, Swendsen_Wang_Physicaa_2016,Puglisi_PhysRep_2017,
Baldovin_JStatMech_2017},
in what has become a lively debate.
Although this may seem a marginal issue, it has crucial consequences about the foundations of statistical mechanics.
For instance the negative temperatures notion wouldn't make sense, since
they are a well founded concept in the Boltzmann description, whereas, they
are forbidden in the case of the Gibbs entropy since the number of microstates
with energy below a given value $E$ is a non-decreasing function of $E$.
Even if we do not share the point of view of authors of Refs.
\cite{Dunkel2013,Hilbert_PRE_2014,Hanggi_15, Campisi_2015,Campisi_2016},
as we have clarified in Refs. \cite{Buonsante_AoP_2016, Buonsante_2015}
where we have shown that the Boltzmann entropy provides a consistent description
of the microcanonical ensemble, in our opinion these authors must be given
credit for having raised this key question.

A further issue raised by the authors of Refs.
\cite{Dunkel2013,Hilbert_PRE_2014,Hanggi_15,
Campisi_2015,Campisi_2016} pertains to the fact that the caloric equation
of state, for instance in the simple case of an isolated ideal gas system,
derived with the Boltzmann entropy is not strictly extensive.
About this point, in Ref. \cite{Buonsante_AoP_2016,Swendsen_2017}, we have shown that the correction to the extensive behaviour, is of the order of $1/(nd)$, and therefore it vanishes in the limit of infinite degrees of freedom.
Although in the case of a macroscopic system (as the ones more often considered in
statistical mechanics) this is not an issue and it represents just an aesthetical
mathematical problem, it pose a relevant  matter when microcanonical thermodynamics
is applied to systems that for their nature do not admit the thermodynamic limit.
Examples of the latter class include proteins, DNA helix, nanosystems.

In the present manuscript we propose a modified version of the Boltzmann
entropy that overcomes all of these issues. In fact, this entropy reproduces
the same results as the Boltzmann entropy for systems with a macroscopic
number of particles and predicts the correct extensivity for the caloric
equation in the case of small systems.
Let $H(x)$ be a classical Hamiltonian describing an autonomous many-body system
of $n$ interacting particles in $d$ spatial dimensions, whose
coordinates and canonical momenta $(q_1\ldots, p_1 ,\ldots)$ are represented as 
$N$-component vectors $x\in \mathbb{R}^{N}$, with $N=2nd$.
Moreover, we assume that no other conserved
quantities do exist in addition to the total energy $H$ \cite{Franzosi_JSP11,Franzosi_PRE12}.
Let 
$
M_E = \left\{x\in \mathbb{R}^{N} | H(x) \leq E \right\}
$
be the set of phase-space states with total energy less than or equal to $E$.
The Gibbs entropy for this system is
\begin{equation}
S_G (E) = \kappa_B \ln \Omega(E) \, ,
\label{gibbs}
\end{equation}
where $\kappa_B$ is the Boltzmann constant and
\begin{equation}
\Omega(E) = \dfrac{1}{h^{nd}} \int d^N x \Theta(E-H(x)) \, ,
\label{OmegaE}
\end{equation}
is the number of states with energy below $E$.  $h$ is the Planck constant and
$\Theta$ is the Heaviside function.

The Boltzmann entropy concerns the energy level sets 
$
\Sigma_E = \left\{x\in \mathbb{R}^{N} | H(x) = E \right\} \, ,
$
and is given in terms of $\omega(E) = \partial \Omega/\partial E$, according to
\begin{equation}
S_B (E) = \kappa_B \ln \left(\omega(E)\Delta \right) \, ,
\label{boltzmann}
\end{equation}
where the constant $\Delta$ with the dimension of energy makes the argument of the logarithm dimensionless,
and
\begin{equation}
\omega(E) = \dfrac{1}{h^{nd}} \int d^N x \delta(E-H(x)) \, ,
\label{omegaE}
\end{equation}
is expressed in terms of the Dirac $\delta$ function. Remarkably, in the case
of smooth level sets $\Sigma_E$, $\omega(E)$
can be cast in the following form 
\cite{RughPRL97,Franzosi_JSP11,Franzosi_PRE12}
\begin{equation}
\omega(E) =  \dfrac{1}{h^{nd}} 
\int_{\Sigma_E} \dfrac{m^{N-1}(\Sigma_E)}{\Vert\nabla H(x) \Vert}  \, ,
\label{omegaEdiff}
\end{equation}
where $m^{N-1}(\Sigma_E)$ is the metric induced from $\mathbb{R}^N$ on the
hypersurface $\Sigma_E$ and $\Vert\nabla H(x) \Vert$ is the norm of the gradient
of $H$ at $x$. 

The entropy that we propose here is
\begin{equation}
S (E) = \kappa_B \ln \left( \sigma(E) \Delta^{1/2} \right) \, ,
\label{enew}
\end{equation}
where
\begin{equation}
\sigma(E) =\dfrac{1}{h^{nd}}  \int_{\Sigma_E} m^{N-1}(\Sigma_E)  \, .
\label{sigmaEdiff}
\end{equation}
In the case of a system of identical particles, to avoid the Gibbs
paradox it is in order to introduce a factor 1/n! in the definitions of $\Omega$,
$\omega$ and $\sigma$, Eqs. \eqref{OmegaE}, \eqref{omegaE}, \eqref{omegaEdiff}
and \eqref{sigmaEdiff}, as we will do in the following.

The entropy is the fundamental thermodynamic potential of the microcanonical ensemble from which
secondary thermodynamic quantities are obtained by derivatives with respect to the control
parameter: the total energy $E$, the occupied volume $V$ and, possibly, further Hamiltonian
parameters $A_\mu$ (in the following we omit to indicate explicitly the dependence by $A_\mu$ in 
order to simplify the notation).
The inverse temperatures $\beta=(\kappa_B T)^{-1}$ is derived from the
the entropy according to $\beta = (\partial S/\partial E)/\kappa_B$, thus in the
three cases under consideration we have
\begin{eqnarray}
\beta_G &=& 
\dfrac{\Omega^\prime}{\Omega} \, , \\
\beta_B &=& 
\dfrac{\omega^\prime}{\omega} \, , \\
\beta &=& 
\dfrac{\sigma^\prime}{\sigma} \, ,
\end{eqnarray}
where the symbol $^\prime$ denotes the partial derivative of the corresponding
term with respect to energy $E$.

A basic requisite for $S$ is to allow the measure of temperature 
and the other secondary
thermodynamic quantities via microcanonical averages.
In terms of the microscopic dynamics, from the Liouville theorem it follows that the
invariant measure $d \mu$ for the dynamics on each energy level-set $\Sigma_E$ is
$d \mu = {m^{N-1}(\Sigma_E)}/{\Vert \nabla H \Vert}$.
In the case of the Boltzmann entropy, the temperature definition meets the
mentioned requisite since
\begin{equation}
\beta_B =\left\langle 
\nabla \left( \frac{\nabla H}{\Vert \nabla H \Vert^2} \right) \right\rangle \, ,
\label{betaB}
\end{equation}
where $\langle \rangle$ indicates the microcanonical average
\begin{equation}
\langle \phi \rangle = \dfrac{1}{\omega} \int_{\Sigma_E} \phi d\mu \, .
\end{equation} 
Eq. \eqref{betaB} is derived in Ref. \cite{RughPRL97}
for the case of many-particle systems for which the energy is the only conserved quantity, and in Refs. 
\cite{Franzosi_JSP11,Franzosi_PRE12} for the general case of two or more
conserved quantities. On the contrary, the Gibbs definition of 
temperature does not meet such important requisite as diffusely
discussed in Ref. \cite{Buonsante_AoP_2016}.
By using by the Federer-Laurence derivation formula \cite{Federer_1969,Laurence_1989,Franzosi_JSP11,Franzosi_PRE12}, 
in the case of the proposed entropy we get
\begin{equation}
\beta = \dfrac{\sigma^\prime}{\sigma} =  \dfrac{\sigma^\prime/\omega}{\sigma/\omega} =
\dfrac{\langle \nabla \left( \frac{\nabla H}{\Vert \nabla H \Vert} \right) \rangle}
{\langle \Vert \nabla H \Vert \rangle} \, .
\label{beta}
\end{equation}
This shows that also $S$, besides $S_B$, satisfies the requirement to provide secondary thermodynamic
quantities measurable as microcanonical averages. In passing, we note that
under the hypothesis of ergodicity, the
averages of each dynamical observable of the system can be equivalently measured along the dynamics.

As a simple test let us consider a classical ideal gas in $d$-spatial
dimensions composed of $n$ 
identical particles of mass $m$ for which it is easy matter to verify that
\begin{eqnarray}
\Omega (E,V) &=& \dfrac{V^n (2\pi m)^{nd/2}}{\Gamma(\frac{nd}{2} +1) n! h^{nd} } E^{nd/2} \, ,
\\
\omega (E,V) &=& \dfrac{V^n (2\pi m)^{nd/2}}{\Gamma(\frac{nd}{2}) n! h^{nd} } E^{nd/2-1} \, ,
\\
\sigma (E,V) &=& \dfrac{2 V^n (2\pi m)^{nd/2}}{\Gamma(\frac{nd}{2}) n! h^{nd} } E^{(nd-1)/2} \, ,
\end{eqnarray}
where the factor $1/n!$ is introduced in order to avoid the Gibbs paradox.
From these formulas one finds the following expression of the caloric 
equation
\begin{eqnarray}
\beta^{-1}_G &=& \dfrac{E}{{nd}/{2}}  \, ,\\
 \beta^{-1}_B &=& \dfrac{E}{\left({nd}/{2}-1\right)} \, , \\
\beta^{-1} &=& \dfrac{E}{{(nd-1)}/{2}} \, .
\end{eqnarray}
In the count for the degrees of freedom for a system of free particles,
just the kinetic term contributes. Thus, in $d$ spatial dimensions a
system of $n$ particles  have $nd$ degrees of freedom and, by setting
the energy  $E$ to a given value we are left with $nd-1$ degrees of
freedom.
Therefore, among these only the latter expression is exactly extensive
and, hence, rigorously satisfies the equipartition theorem for any $n$.
With an analogous calculation, it is easy matter to show that for a system
of $n$ independent identical harmonic oscillators, of mass $m$ and
frequency $\nu$ in $d$ spatial dimensions the caloric equations
derived from the three entropies are
\begin{eqnarray}
\beta^{-1}_G &=& \dfrac{E}{nd}  \, ,\\ 
\beta^{-1}_B &=& \dfrac{E}{\left({nd}-1\right)} \, , \\
\beta^{-1} &=& \dfrac{E}{{(2nd-1)}/{2}}  \, .
\end{eqnarray}
In this case, either the coordinates and the motional degrees of freedom
contribute to the count of the degrees of freedom of the system.
Thus, when the energy has a fixed value $E$, the number of degrees
of freedom are $2nd-1$ and only $S$ brings to the correct equipartition
formula.

\emph{In addition to lead up the correct relation between total energy
and true number of degrees of freedom, the entropy we propose rigorously
satisfies the postulate of equal a-priory probability which is a very
foundations of the equilibrium microcanonic statistical mechanics.
As a matter of fact, for a generic isolated physical-system at the
equilibrium, a given thermodynamic state is completely determined when
we know the values of the macroscopic parameters as energy, volume, and
possibly further external parameters, that characterize such system.
In this way, from a thermodynamic point of view we do not distinguish
between the states of the system represented by different points on
the same energy level and consistent with the further constraints.
This is just what Eq. \eqref{omegaEdiff} does, it ``counts the number´´
of microstates satisfying the macroscopic constraint $H=E$,
consistently to the above mentioned postulate.
On the contrary, the standard
Boltzmann entropy adopts a place-dependent weight
$1/\Vert \nabla H \Vert$.}

In order to better clarify the connection between the Boltzmann
entropy and that one we propose, let us perform the following
rough calculation.
For a system with $N$ degrees of freedom, if $\Delta E \ll E$,
approximatively we have
\begin{equation}
\Omega(E+\Delta E) - \Omega(E) \approx \omega(E) \Delta E + O(\Delta E^{2})\, ,
\end{equation}
on the other hand, for the Cavalieri's principle, we have
\begin{equation}
\Omega(E+\Delta E ) - \Omega(E) \approx 
\left(\sigma(E)\Delta^{1/2} 
\right) \dfrac{\Delta E}{\Delta}  + O(\Delta E^{2})
\, .
\end{equation}
Hence it results
$
\sigma (E)\Delta^{1/2} = \omega(E)\Delta + O(N^{2})
$
and, consequently
\begin{equation}
\lim_{N\to  \infty} \dfrac{1}{N} \left(\ln (\sigma \Delta^{1/2})-\ln(\omega \Delta)
\right)
= 0 \, .
\label{bigN}
\end{equation}
This makes evident that in the limit of large number of degrees of freedom,
the proposed entropy predicts the same results as the Boltzmann entropy,
whereas, in the case of systems with small $N$ the two
entropies differ from each other.

In order to verify our assumption, we have tested the proposed
entropy on two systems: the two dimensional $\Phi^4$ model and
a one dimensional model of rotors.

The $\phi^4$ model \cite{Franzosi_PRE99,Franzosi_PRL00,BPV_PRB04,Franzosi_PRA10} is defined by the Hamiltonian
\begin{equation}
H = \sum_{\bf j} \dfrac{1}{2} \pi^2_{\bf j}  + V(\phi)
\label{Hphi4}
\end{equation}
where
\begin{equation}
V(\phi) =\sum_{\bf j} \left[ 
 \dfrac{\lambda}{4!} \phi^4_{\bf j} - \dfrac{\mu^2}{2}\phi^2_{\bf j} +
\dfrac{J}{4} \sum_{{\bf k}\in I({\bf j})} (\phi_{\bf j} - \phi_{\bf k})^2
 \right] \, ,
\label{Vphi4}
\end{equation}
$\pi_{\bf j}$ is the conjugate momentum of the variable $\phi_{\bf j}$
that defines the field at ${\bf j}^{th}$ site. Indeed,
${\bf j} = (j_1,j_2)$ denotes a site of a two dimensional latte
and
$I({\bf j})$ are the nearest neighbour lattice sites of the
${\bf j}^{th}$ site. The coordinates of the sites are integer numbers
$j_k =1,\ldots,N_k$, $k=1,2$, so that the total number of sites
in the lattice is $N=N_1\,N_2$. Furthermore periodic boundary conditions
are assumed.
The local potential displays a double-well shape whose minima are located 
at $\pm \sqrt{{3! \mu^2}/{\lambda}}$ and to which it corresponds 
the ground-state energy per particle $e_0 = - 3! \mu^4/(2 \lambda)$.
At low-energies the system is dominated by an ordered phase where the time 
averages of the local field are not vanishing. By increasing the system
energy the system undergoes a second order phase-transition and
the local $\mathbb{Z}_2$ symmetry is restored. In fact, at high
energies the time averages of the local field go to zero.

The second model \cite{Cerino_2015} is composed by $N$ rotators with
canonical coordinates
$\phi_1,\ldots,\phi_N,\pi_1,\ldots,\pi_N$ and Hamiltonian
\begin{equation}
H=\sum^N_{j=1} [1-\cos(\pi_j)] + \epsilon 
\sum^N_{j=1}[1-\cos(\phi_j-\phi_{j-1})] \, ,
\label{Hrot}
\end{equation}
where is assumed $\phi_0 = 0$. The form of kinetic and potential terms
in \eqref{Hrot} makes the energy bounded either from above and from below
and such Hamiltonian implies the existence of negative Boltzmann temperatures
\cite{Cerino_2015}.

We have numerically integrated the equation of motion associated to the 
Hamiltonian of both the models, by using a third order symplectic algorithm
and starting from initial conditions corresponding to different values
of the system total energy $E$. We have measured along the dynamics the time averages of the relevant quantities that appear in \eqref{betaB} 
and \eqref{beta} and, then we have derived the curves $\beta_B(E)$ and
$\beta(E)$ for the two models.
\begin{figure}[h]
 \includegraphics[height=5.cm]{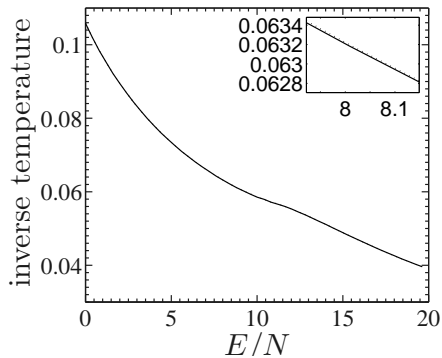}
\caption{The figure compares $\beta_B(E/N)$ (dotted line) and
$\beta(E/N)$ (continuous line) numerically computed for a lattice
of $128\times 128$ sites for the
$\Phi^4$-model. The agreement is astonishing,
in fact the two curves are indistinguishable. In the inset we report
a zoom in order to show the two curves.
\label{fig1}}
\end{figure} 
\begin{figure}[h]
 \includegraphics[height=5.cm]{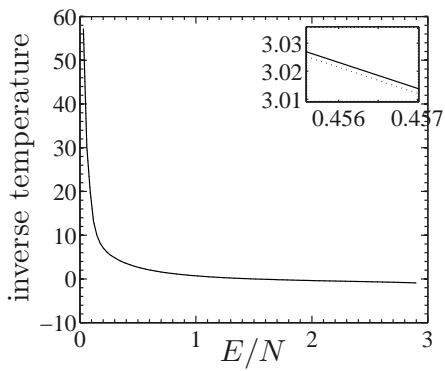}
\caption{The figure compares $\beta_B(E/N)$ (dotted line) and
$\beta(E/N)$ (continuous line) numerically computed for an array of $512$
rotors. Also here the agreement is astonishing,
the two curves are indistinguishable thus we report the inset with
a zoom that shows the two curves.
\label{fig2}}
\end{figure} 
Figs. 1 and 2 clearly show the remarkable agreement between the curves
$\beta_B(E/N)$ and $\beta(E/N)$, for both the models studied.

In conclusion we have proposed a novel definition of the microcanonical
entropy for classical systems.
We have shown that this definition definitely resolve the debate
on the correct definition of the microcanonical entropy. In fact, we have
shown that this entropy definition fixes the issue inherent the full
extensivity of the caloric equation.
Furthermore, we have given evidence by investigating
two different models, that this entropy reproduces results which are in
agreement with the ones predicted with standard Boltzmann entropy in the
case of macroscopic systems. 
Since the differences between the predictions of Boltzmann entropy and
of the one here proposed, are more evident in systems with small number of
degrees of freedom,  we conclude that the Boltzmann entropy (with the our one)
provides a correct description for macroscopic systems whereas
extremely small systems should be described with the entropy that we
have proposed in order to avoid, for instance, issues with the
extensivity of the caloric equation.

\begin{acknowledgments}
We are grateful to A. Smerzi and P. Buonsante for useful discussions.
\end{acknowledgments}

\bibliography{references}

\begin{thebibliography}{47}%
\makeatletter
\providecommand \@ifxundefined [1]{%
 \@ifx{#1\undefined}
}%
\providecommand \@ifnum [1]{%
 \ifnum #1\expandafter \@firstoftwo
 \else \expandafter \@secondoftwo
 \fi
}%
\providecommand \@ifx [1]{%
 \ifx #1\expandafter \@firstoftwo
 \else \expandafter \@secondoftwo
 \fi
}%
\providecommand \natexlab [1]{#1}%
\providecommand \enquote  [1]{``#1''}%
\providecommand \bibnamefont  [1]{#1}%
\providecommand \bibfnamefont [1]{#1}%
\providecommand \citenamefont [1]{#1}%
\providecommand \href@noop [0]{\@secondoftwo}%
\providecommand \href [0]{\begingroup \@sanitize@url \@href}%
\providecommand \@href[1]{\@@startlink{#1}\@@href}%
\providecommand \@@href[1]{\endgroup#1\@@endlink}%
\providecommand \@sanitize@url [0]{\catcode `\\12\catcode `\$12\catcode
  `\&12\catcode `\#12\catcode `\^12\catcode `\_12\catcode `\%12\relax}%
\providecommand \@@startlink[1]{}%
\providecommand \@@endlink[0]{}%
\providecommand \url  [0]{\begingroup\@sanitize@url \@url }%
\providecommand \@url [1]{\endgroup\@href {#1}{\urlprefix }}%
\providecommand \urlprefix  [0]{URL }%
\providecommand \Eprint [0]{\href }%
\providecommand \doibase [0]{http://dx.doi.org/}%
\providecommand \selectlanguage [0]{\@gobble}%
\providecommand \bibinfo  [0]{\@secondoftwo}%
\providecommand \bibfield  [0]{\@secondoftwo}%
\providecommand \translation [1]{[#1]}%
\providecommand \BibitemOpen [0]{}%
\providecommand \bibitemStop [0]{}%
\providecommand \bibitemNoStop [0]{.\EOS\space}%
\providecommand \EOS [0]{\spacefactor3000\relax}%
\providecommand \BibitemShut  [1]{\csname bibitem#1\endcsname}%
\let\auto@bib@innerbib\@empty
\bibitem [{\citenamefont {Georgescu-Roegen}(1993)}]{entropy_economic}%
  \BibitemOpen
  \bibfield  {author} {\bibinfo {author} {\bibfnamefont {N.}~\bibnamefont
  {Georgescu-Roegen}},\ }\href@noop {} {\bibfield  {journal} {\bibinfo
  {journal} {Valuing the earth: Economics, ecology, ethics}\ ,\ \bibinfo
  {pages} {75}} (\bibinfo {year} {1993})}\BibitemShut {NoStop}%
\bibitem [{\citenamefont {Tsallis}\ and\ \citenamefont
  {Brigatti}(2004)}]{tsallis2004}%
  \BibitemOpen
  \bibfield  {author} {\bibinfo {author} {\bibfnamefont {C.}~\bibnamefont
  {Tsallis}}\ and\ \bibinfo {author} {\bibfnamefont {E.}~\bibnamefont
  {Brigatti}},\ }\href@noop {} {\bibfield  {journal} {\bibinfo  {journal}
  {Continuum Mechanics and Thermodynamics}\ }\textbf {\bibinfo {volume} {16}},\
  \bibinfo {pages} {223} (\bibinfo {year} {2004})}\BibitemShut {NoStop}%
\bibitem [{\citenamefont {Banavar}\ \emph {et~al.}(2010)\citenamefont
  {Banavar}, \citenamefont {Maritan},\ and\ \citenamefont
  {Volkov}}]{entropy_ecology}%
  \BibitemOpen
  \bibfield  {author} {\bibinfo {author} {\bibfnamefont {J.~R.}\ \bibnamefont
  {Banavar}}, \bibinfo {author} {\bibfnamefont {A.}~\bibnamefont {Maritan}}, \
  and\ \bibinfo {author} {\bibfnamefont {I.}~\bibnamefont {Volkov}},\ }\href
  {http://stacks.iop.org/0953-8984/22/i=6/a=063101} {\bibfield  {journal}
  {\bibinfo  {journal} {Journal of Physics: Condensed Matter}\ }\textbf
  {\bibinfo {volume} {22}},\ \bibinfo {pages} {063101} (\bibinfo {year}
  {2010})}\BibitemShut {NoStop}%
\bibitem [{\citenamefont {Franzosi}\ \emph {et~al.}(2015)\citenamefont
  {Franzosi}, \citenamefont {Felice}, \citenamefont {Mancini},\ and\
  \citenamefont {Pettini}}]{Franzosi_EPL15}%
  \BibitemOpen
  \bibfield  {author} {\bibinfo {author} {\bibfnamefont {R.}~\bibnamefont
  {Franzosi}}, \bibinfo {author} {\bibfnamefont {D.}~\bibnamefont {Felice}},
  \bibinfo {author} {\bibfnamefont {S.}~\bibnamefont {Mancini}}, \ and\
  \bibinfo {author} {\bibfnamefont {M.}~\bibnamefont {Pettini}},\ }\href
  {http://stacks.iop.org/0295-5075/111/i=2/a=20001} {\bibfield  {journal}
  {\bibinfo  {journal} {EPL (Europhysics Letters)}\ }\textbf {\bibinfo {volume}
  {111}},\ \bibinfo {pages} {20001} (\bibinfo {year} {2015})}\BibitemShut
  {NoStop}%
\bibitem [{\citenamefont {Franzosi}\ \emph {et~al.}(2016)\citenamefont
  {Franzosi}, \citenamefont {Felice}, \citenamefont {Mancini},\ and\
  \citenamefont {Pettini}}]{Franzosi_PRE16}%
  \BibitemOpen
  \bibfield  {author} {\bibinfo {author} {\bibfnamefont {R.}~\bibnamefont
  {Franzosi}}, \bibinfo {author} {\bibfnamefont {D.}~\bibnamefont {Felice}},
  \bibinfo {author} {\bibfnamefont {S.}~\bibnamefont {Mancini}}, \ and\
  \bibinfo {author} {\bibfnamefont {M.}~\bibnamefont {Pettini}},\ }\href
  {\doibase 10.1103/PhysRevE.93.062317} {\bibfield  {journal} {\bibinfo
  {journal} {Phys. Rev. E}\ }\textbf {\bibinfo {volume} {93}},\ \bibinfo
  {pages} {062317} (\bibinfo {year} {2016})}\BibitemShut {NoStop}%
\bibitem [{\citenamefont {Felice}\ \emph {et~al.}(2018)\citenamefont {Felice},
  \citenamefont {Franzosi}, \citenamefont {Mancini},\ and\ \citenamefont
  {Pettini}}]{Felice_PhysA_2018}%
  \BibitemOpen
  \bibfield  {author} {\bibinfo {author} {\bibfnamefont {D.}~\bibnamefont
  {Felice}}, \bibinfo {author} {\bibfnamefont {R.}~\bibnamefont {Franzosi}},
  \bibinfo {author} {\bibfnamefont {S.}~\bibnamefont {Mancini}}, \ and\
  \bibinfo {author} {\bibfnamefont {M.}~\bibnamefont {Pettini}},\ }\href
  {\doibase https://doi.org/10.1016/j.physa.2017.09.007} {\bibfield  {journal}
  {\bibinfo  {journal} {Physica A: Statistical Mechanics and its Applications}\
  }\textbf {\bibinfo {volume} {491}},\ \bibinfo {pages} {666 } (\bibinfo {year}
  {2018})}\BibitemShut {NoStop}%
\bibitem [{\citenamefont {Hertz}(1910)}]{Hertz10}%
  \BibitemOpen
  \bibfield  {author} {\bibinfo {author} {\bibfnamefont {P.}~\bibnamefont
  {Hertz}},\ }\href@noop {} {\bibfield  {journal} {\bibinfo  {journal} {Ann.
  Phys. (Leipzig)}\ }\textbf {\bibinfo {volume} {338}},\ \bibinfo {pages} {225}
  (\bibinfo {year} {1910})}\BibitemShut {NoStop}%
\bibitem [{\citenamefont {Einstein}(1911)}]{Einstein11}%
  \BibitemOpen
  \bibfield  {author} {\bibinfo {author} {\bibfnamefont {A.}~\bibnamefont
  {Einstein}},\ }\href@noop {} {\bibfield  {journal} {\bibinfo  {journal} {Ann.
  Phys. (Leipzig)}\ }\textbf {\bibinfo {volume} {339}},\ \bibinfo {pages} {175}
  (\bibinfo {year} {1911})}\BibitemShut {NoStop}%
\bibitem [{\citenamefont {Schl\"uter}(1948)}]{Schluter48}%
  \BibitemOpen
  \bibfield  {author} {\bibinfo {author} {\bibfnamefont {A.}~\bibnamefont
  {Schl\"uter}},\ }\href@noop {} {\bibfield  {journal} {\bibinfo  {journal} {Z.
  Naturforschg.}\ }\textbf {\bibinfo {volume} {3a}},\ \bibinfo {pages} {350}
  (\bibinfo {year} {1948})}\BibitemShut {NoStop}%
\bibitem [{\citenamefont {Jaynes}(1965)}]{Jaynes}%
  \BibitemOpen
  \bibfield  {author} {\bibinfo {author} {\bibfnamefont {E.~T.}\ \bibnamefont
  {Jaynes}},\ }\href {\doibase http://dx.doi.org/10.1119/1.1971557} {\bibfield
  {journal} {\bibinfo  {journal} {American Journal of Physics}\ }\textbf
  {\bibinfo {volume} {33}},\ \bibinfo {pages} {391} (\bibinfo {year}
  {1965})}\BibitemShut {NoStop}%
\bibitem [{\citenamefont {M\"unster}(1969)}]{Munster_1987}%
  \BibitemOpen
  \bibfield  {author} {\bibinfo {author} {\bibfnamefont {A.}~\bibnamefont
  {M\"unster}},\ }\href@noop {} {\emph {\bibinfo {title} {Statistical
  Thermodynamics}}}\ (\bibinfo  {publisher} {Springer Berlin},\ \bibinfo {year}
  {1969})\BibitemShut {NoStop}%
\bibitem [{\citenamefont {Pearson}\ \emph {et~al.}(1985)\citenamefont
  {Pearson}, \citenamefont {Halicioglu},\ and\ \citenamefont
  {Tiller}}]{Pearson85}%
  \BibitemOpen
  \bibfield  {author} {\bibinfo {author} {\bibfnamefont {E.~M.}\ \bibnamefont
  {Pearson}}, \bibinfo {author} {\bibfnamefont {T.}~\bibnamefont {Halicioglu}},
  \ and\ \bibinfo {author} {\bibfnamefont {W.~A.}\ \bibnamefont {Tiller}},\
  }\href {\doibase 10.1103/PhysRevA.32.3030} {\bibfield  {journal} {\bibinfo
  {journal} {Phys. Rev. A}\ }\textbf {\bibinfo {volume} {32}},\ \bibinfo
  {pages} {3030} (\bibinfo {year} {1985})}\BibitemShut {NoStop}%
\bibitem [{\citenamefont {Berdichevsky}\ \emph {et~al.}(1991)\citenamefont
  {Berdichevsky}, \citenamefont {Kunin},\ and\ \citenamefont
  {Hussain}}]{Berdichevsky91}%
  \BibitemOpen
  \bibfield  {author} {\bibinfo {author} {\bibfnamefont {V.}~\bibnamefont
  {Berdichevsky}}, \bibinfo {author} {\bibfnamefont {I.}~\bibnamefont {Kunin}},
  \ and\ \bibinfo {author} {\bibfnamefont {F.}~\bibnamefont {Hussain}},\ }\href
  {\doibase 10.1103/PhysRevA.43.2050} {\bibfield  {journal} {\bibinfo
  {journal} {Phys. Rev. A}\ }\textbf {\bibinfo {volume} {43}},\ \bibinfo
  {pages} {2050} (\bibinfo {year} {1991})}\BibitemShut {NoStop}%
\bibitem [{\citenamefont {Adib}(2004)}]{Adib04}%
  \BibitemOpen
  \bibfield  {author} {\bibinfo {author} {\bibfnamefont {A.}~\bibnamefont
  {Adib}},\ }\href@noop {} {\bibfield  {journal} {\bibinfo  {journal} {Journal
  of Statistical Physics}\ }\textbf {\bibinfo {volume} {117}},\ \bibinfo
  {pages} {581} (\bibinfo {year} {2004})}\BibitemShut {NoStop}%
\bibitem [{\citenamefont {Lavis}(2005)}]{Lavis2005245}%
  \BibitemOpen
  \bibfield  {author} {\bibinfo {author} {\bibfnamefont {D.}~\bibnamefont
  {Lavis}},\ }\href {\doibase http://dx.doi.org/10.1016/j.shpsb.2004.11.007}
  {\bibfield  {journal} {\bibinfo  {journal} {Studies in History and Philosophy
  of Science Part B: Studies in History and Philosophy of Modern Physics}\
  }\textbf {\bibinfo {volume} {36}},\ \bibinfo {pages} {245 } (\bibinfo {year}
  {2005})}\BibitemShut {NoStop}%
\bibitem [{\citenamefont {Campisi}(2005)}]{Campisi05}%
  \BibitemOpen
  \bibfield  {author} {\bibinfo {author} {\bibfnamefont {M.}~\bibnamefont
  {Campisi}},\ }\href {\doibase http://dx.doi.org/10.1016/j.shpsb.2005.01.001}
  {\bibfield  {journal} {\bibinfo  {journal} {Studies in History and Philosophy
  of Science Part B: Studies in History and Philosophy of Modern Physics}\
  }\textbf {\bibinfo {volume} {36}},\ \bibinfo {pages} {275 } (\bibinfo {year}
  {2005})}\BibitemShut {NoStop}%
\bibitem [{\citenamefont {Dunkel}\ and\ \citenamefont
  {Hilbert}(2013)}]{Dunkel2013}%
  \BibitemOpen
  \bibfield  {author} {\bibinfo {author} {\bibfnamefont {J.}~\bibnamefont
  {Dunkel}}\ and\ \bibinfo {author} {\bibfnamefont {S.}~\bibnamefont
  {Hilbert}},\ }\href {\doibase 10.1038/nphys2815} {\bibfield  {journal}
  {\bibinfo  {journal} {Nature Physics}\ }\textbf {\bibinfo {volume} {10}},\
  \bibinfo {pages} {67} (\bibinfo {year} {2013})}\BibitemShut {NoStop}%
\bibitem [{\citenamefont {Hilbert}\ \emph {et~al.}(2014)\citenamefont
  {Hilbert}, \citenamefont {H\"anggi},\ and\ \citenamefont
  {Dunkel}}]{Hilbert_PRE_2014}%
  \BibitemOpen
  \bibfield  {author} {\bibinfo {author} {\bibfnamefont {S.}~\bibnamefont
  {Hilbert}}, \bibinfo {author} {\bibfnamefont {P.}~\bibnamefont {H\"anggi}}, \
  and\ \bibinfo {author} {\bibfnamefont {J.}~\bibnamefont {Dunkel}},\ }\href
  {\doibase 10.1103/PhysRevE.90.062116} {\bibfield  {journal} {\bibinfo
  {journal} {Phys. Rev. E}\ }\textbf {\bibinfo {volume} {90}},\ \bibinfo
  {pages} {062116} (\bibinfo {year} {2014})}\BibitemShut {NoStop}%
\bibitem [{\citenamefont {H\"anggi}\ \emph {et~al.}(2015)\citenamefont
  {H\"anggi}, \citenamefont {Hilbert},\ and\ \citenamefont
  {Dunkel}}]{Hanggi_15}%
  \BibitemOpen
  \bibfield  {author} {\bibinfo {author} {\bibfnamefont {P.}~\bibnamefont
  {H\"anggi}}, \bibinfo {author} {\bibfnamefont {S.}~\bibnamefont {Hilbert}}, \
  and\ \bibinfo {author} {\bibfnamefont {J.}~\bibnamefont {Dunkel}},\
  }\href@noop {} {\bibfield  {journal} {\bibinfo  {journal} {arXiv:1507.05713}\
  } (\bibinfo {year} {2015})}\BibitemShut {NoStop}%
\bibitem [{\citenamefont {Sokolov}(2014)}]{Sokolov_2014}%
  \BibitemOpen
  \bibfield  {author} {\bibinfo {author} {\bibfnamefont {I.~M.}\ \bibnamefont
  {Sokolov}},\ }\href {http://dx.doi.org/10.1038/nphys2831} {\bibfield
  {journal} {\bibinfo  {journal} {Nat Phys}\ }\textbf {\bibinfo {volume}
  {10}},\ \bibinfo {pages} {7} (\bibinfo {year} {2014})}\BibitemShut {NoStop}%
\bibitem [{\citenamefont {Dunkel}\ and\ \citenamefont
  {Hilbert}(2014{\natexlab{a}})}]{DunkelHilbertRep1}%
  \BibitemOpen
  \bibfield  {author} {\bibinfo {author} {\bibfnamefont {J.}~\bibnamefont
  {Dunkel}}\ and\ \bibinfo {author} {\bibfnamefont {S.}~\bibnamefont
  {Hilbert}},\ }\href@noop {} {\bibfield  {journal} {\bibinfo  {journal}
  {arXiv:1403.6058}\ } (\bibinfo {year} {2014}{\natexlab{a}})}\BibitemShut
  {NoStop}%
\bibitem [{\citenamefont {Dunkel}\ and\ \citenamefont
  {Hilbert}(2014{\natexlab{b}})}]{DunkelHilbertRep2}%
  \BibitemOpen
  \bibfield  {author} {\bibinfo {author} {\bibfnamefont {J.}~\bibnamefont
  {Dunkel}}\ and\ \bibinfo {author} {\bibfnamefont {S.}~\bibnamefont
  {Hilbert}},\ }\href@noop {} {\bibfield  {journal} {\bibinfo  {journal}
  {arXiv:1408.5392}\ } (\bibinfo {year} {2014}{\natexlab{b}})}\BibitemShut
  {NoStop}%
\bibitem [{\citenamefont {Campisi}(2015)}]{Campisi_2015}%
  \BibitemOpen
  \bibfield  {author} {\bibinfo {author} {\bibfnamefont {M.}~\bibnamefont
  {Campisi}},\ }\href {\doibase 10.1103/PhysRevE.91.052147} {\bibfield
  {journal} {\bibinfo  {journal} {Phys. Rev. E}\ }\textbf {\bibinfo {volume}
  {91}},\ \bibinfo {pages} {052147} (\bibinfo {year} {2015})}\BibitemShut
  {NoStop}%
\bibitem [{\citenamefont {Campisi}(2016)}]{Campisi_2016}%
  \BibitemOpen
  \bibfield  {author} {\bibinfo {author} {\bibfnamefont {M.}~\bibnamefont
  {Campisi}},\ }\href {\doibase 10.1103/PhysRevE.93.039901} {\bibfield
  {journal} {\bibinfo  {journal} {Phys. Rev. E}\ }\textbf {\bibinfo {volume}
  {93}},\ \bibinfo {pages} {039901} (\bibinfo {year} {2016})}\BibitemShut
  {NoStop}%
\bibitem [{\citenamefont {Romero-Roch\'{i}n}(2013)}]{Romero-Rochin}%
  \BibitemOpen
  \bibfield  {author} {\bibinfo {author} {\bibfnamefont {V.}~\bibnamefont
  {Romero-Roch\'{i}n}},\ }\href {\doibase 10.1103/PhysRevE.88.022144}
  {\bibfield  {journal} {\bibinfo  {journal} {Phys. Rev. E}\ }\textbf {\bibinfo
  {volume} {88}},\ \bibinfo {pages} {022144} (\bibinfo {year}
  {2013})}\BibitemShut {NoStop}%
\bibitem [{\citenamefont {Treumann}\ and\ \citenamefont
  {Baumjohann}(2014{\natexlab{a}})}]{Treumann_2014}%
  \BibitemOpen
  \bibfield  {author} {\bibinfo {author} {\bibfnamefont {R.~A.}\ \bibnamefont
  {Treumann}}\ and\ \bibinfo {author} {\bibfnamefont {W.}~\bibnamefont
  {Baumjohann}},\ }\href@noop {} {\bibfield  {journal} {\bibinfo  {journal}
  {arXiv:1406.6639}\ } (\bibinfo {year} {2014}{\natexlab{a}})}\BibitemShut
  {NoStop}%
\bibitem [{\citenamefont {Treumann}\ and\ \citenamefont
  {Baumjohann}(2014{\natexlab{b}})}]{Treumann_2014a}%
  \BibitemOpen
  \bibfield  {author} {\bibinfo {author} {\bibfnamefont {R.~A.}\ \bibnamefont
  {Treumann}}\ and\ \bibinfo {author} {\bibfnamefont {W.}~\bibnamefont
  {Baumjohann}},\ }\href {\doibase 10.3389/fphy.2014.00049} {\bibfield
  {journal} {\bibinfo  {journal} {Frontiers in Physics}\ }\textbf {\bibinfo
  {volume} {2}} (\bibinfo {year} {2014}{\natexlab{b}}),\
  10.3389/fphy.2014.00049}\BibitemShut {NoStop}%
\bibitem [{\citenamefont {Vilar}\ and\ \citenamefont
  {Rubi}(2014)}]{Vilar_2014}%
  \BibitemOpen
  \bibfield  {author} {\bibinfo {author} {\bibfnamefont {J.~M.~G.}\
  \bibnamefont {Vilar}}\ and\ \bibinfo {author} {\bibfnamefont {J.~M.}\
  \bibnamefont {Rubi}},\ }\href {\doibase http://dx.doi.org/10.1063/1.4879553}
  {\bibfield  {journal} {\bibinfo  {journal} {The Journal of Chemical Physics}\
  }\textbf {\bibinfo {volume} {140}},\ \bibinfo {eid} {201101} (\bibinfo {year}
  {2014})}\BibitemShut {NoStop}%
\bibitem [{\citenamefont {Frenkel}\ and\ \citenamefont
  {Warren}(2015)}]{Frenkel_2015}%
  \BibitemOpen
  \bibfield  {author} {\bibinfo {author} {\bibfnamefont {D.}~\bibnamefont
  {Frenkel}}\ and\ \bibinfo {author} {\bibfnamefont {P.~B.}\ \bibnamefont
  {Warren}},\ }\href {\doibase http://dx.doi.org/10.1119/1.4895828} {\bibfield
  {journal} {\bibinfo  {journal} {American Journal of Physics}\ }\textbf
  {\bibinfo {volume} {83}},\ \bibinfo {pages} {163} (\bibinfo {year}
  {2015})}\BibitemShut {NoStop}%
\bibitem [{\citenamefont {Schneider}\ \emph {et~al.}(2014)\citenamefont
  {Schneider}, \citenamefont {Mandt}, \citenamefont {Rapp}, \citenamefont
  {Braun}, \citenamefont {Weimer}, \citenamefont {Bloch},\ and\ \citenamefont
  {Rosch}}]{Schneider_2014}%
  \BibitemOpen
  \bibfield  {author} {\bibinfo {author} {\bibfnamefont {U.}~\bibnamefont
  {Schneider}}, \bibinfo {author} {\bibfnamefont {S.}~\bibnamefont {Mandt}},
  \bibinfo {author} {\bibfnamefont {A.}~\bibnamefont {Rapp}}, \bibinfo {author}
  {\bibfnamefont {S.}~\bibnamefont {Braun}}, \bibinfo {author} {\bibfnamefont
  {H.}~\bibnamefont {Weimer}}, \bibinfo {author} {\bibfnamefont
  {I.}~\bibnamefont {Bloch}}, \ and\ \bibinfo {author} {\bibfnamefont
  {A.}~\bibnamefont {Rosch}},\ }\href@noop {} {\bibfield  {journal} {\bibinfo
  {journal} {arXiv:1407.4127}\ } (\bibinfo {year} {2014})}\BibitemShut
  {NoStop}%
\bibitem [{\citenamefont {Wang}(2015)}]{Wang_2015}%
  \BibitemOpen
  \bibfield  {author} {\bibinfo {author} {\bibfnamefont {J.-S.}\ \bibnamefont
  {Wang}},\ }\href@noop {} {\bibfield  {journal} {\bibinfo  {journal}
  {arXiv:1507.02022}\ } (\bibinfo {year} {2015})}\BibitemShut {NoStop}%
\bibitem [{\citenamefont {Cerino}\ \emph {et~al.}(2015)\citenamefont {Cerino},
  \citenamefont {Puglisi},\ and\ \citenamefont {Vulpiani}}]{Cerino_2015}%
  \BibitemOpen
  \bibfield  {author} {\bibinfo {author} {\bibfnamefont {L.}~\bibnamefont
  {Cerino}}, \bibinfo {author} {\bibfnamefont {A.}~\bibnamefont {Puglisi}}, \
  and\ \bibinfo {author} {\bibfnamefont {A.}~\bibnamefont {Vulpiani}},\ }\href
  {http://stacks.iop.org/1742-5468/2015/i=12/a=P12002} {\bibfield  {journal}
  {\bibinfo  {journal} {Journal of Statistical Mechanics: Theory and
  Experiment}\ }\textbf {\bibinfo {volume} {2015}},\ \bibinfo {pages} {P12002}
  (\bibinfo {year} {2015})}\BibitemShut {NoStop}%
\bibitem [{\citenamefont {Swendsen}\ and\ \citenamefont
  {Wang}(2016)}]{Swendsen_Wang_Physicaa_2016}%
  \BibitemOpen
  \bibfield  {author} {\bibinfo {author} {\bibfnamefont {R.~H.}\ \bibnamefont
  {Swendsen}}\ and\ \bibinfo {author} {\bibfnamefont {J.-S.}\ \bibnamefont
  {Wang}},\ }\href {\doibase http://dx.doi.org/10.1016/j.physa.2016.01.068}
  {\bibfield  {journal} {\bibinfo  {journal} {Physica A: Statistical Mechanics
  and its Applications}\ }\textbf {\bibinfo {volume} {453}},\ \bibinfo {pages}
  {24 } (\bibinfo {year} {2016})}\BibitemShut {NoStop}%
\bibitem [{\citenamefont {Puglisi}\ \emph {et~al.}(2017)\citenamefont
  {Puglisi}, \citenamefont {Sarracino},\ and\ \citenamefont
  {Vulpiani}}]{Puglisi_PhysRep_2017}%
  \BibitemOpen
  \bibfield  {author} {\bibinfo {author} {\bibfnamefont {A.}~\bibnamefont
  {Puglisi}}, \bibinfo {author} {\bibfnamefont {A.}~\bibnamefont {Sarracino}},
  \ and\ \bibinfo {author} {\bibfnamefont {A.}~\bibnamefont {Vulpiani}},\
  }\href {\doibase https://doi.org/10.1016/j.physrep.2017.09.001} {\bibfield
  {journal} {\bibinfo  {journal} {Physics Reports}\ }\textbf {\bibinfo {volume}
  {709-710}},\ \bibinfo {pages} {1 } (\bibinfo {year} {2017})},\ \bibinfo
  {note} {temperature in and out of equilibrium: a review of concepts, tools
  and attempts}\BibitemShut {NoStop}%
\bibitem [{\citenamefont {Baldovin}\ \emph {et~al.}(2017)\citenamefont
  {Baldovin}, \citenamefont {Puglisi}, \citenamefont {Sarracino},\ and\
  \citenamefont {Vulpiani}}]{Baldovin_JStatMech_2017}%
  \BibitemOpen
  \bibfield  {author} {\bibinfo {author} {\bibfnamefont {M.}~\bibnamefont
  {Baldovin}}, \bibinfo {author} {\bibfnamefont {A.}~\bibnamefont {Puglisi}},
  \bibinfo {author} {\bibfnamefont {A.}~\bibnamefont {Sarracino}}, \ and\
  \bibinfo {author} {\bibfnamefont {A.}~\bibnamefont {Vulpiani}},\ }\href
  {http://stacks.iop.org/1742-5468/2017/i=11/a=113202} {\bibfield  {journal}
  {\bibinfo  {journal} {Journal of Statistical Mechanics: Theory and
  Experiment}\ }\textbf {\bibinfo {volume} {2017}},\ \bibinfo {pages} {113202}
  (\bibinfo {year} {2017})}\BibitemShut {NoStop}%
\bibitem [{\citenamefont {Buonsante}\ \emph {et~al.}(2016)\citenamefont
  {Buonsante}, \citenamefont {Franzosi},\ and\ \citenamefont
  {Smerzi}}]{Buonsante_AoP_2016}%
  \BibitemOpen
  \bibfield  {author} {\bibinfo {author} {\bibfnamefont {P.}~\bibnamefont
  {Buonsante}}, \bibinfo {author} {\bibfnamefont {R.}~\bibnamefont {Franzosi}},
  \ and\ \bibinfo {author} {\bibfnamefont {A.}~\bibnamefont {Smerzi}},\ }\href
  {\doibase http://dx.doi.org/10.1016/j.aop.2016.10.017} {\bibfield  {journal}
  {\bibinfo  {journal} {Annals of Physics}\ }\textbf {\bibinfo {volume}
  {375}},\ \bibinfo {pages} {414 } (\bibinfo {year} {2016})}\BibitemShut
  {NoStop}%
\bibitem [{\citenamefont {Buonsante}\ \emph {et~al.}(2017)\citenamefont
  {Buonsante}, \citenamefont {Franzosi},\ and\ \citenamefont
  {Smerzi}}]{Buonsante_2015}%
  \BibitemOpen
  \bibfield  {author} {\bibinfo {author} {\bibfnamefont {P.}~\bibnamefont
  {Buonsante}}, \bibinfo {author} {\bibfnamefont {R.}~\bibnamefont {Franzosi}},
  \ and\ \bibinfo {author} {\bibfnamefont {A.}~\bibnamefont {Smerzi}},\ }\href
  {\doibase 10.1103/PhysRevE.95.052135} {\bibfield  {journal} {\bibinfo
  {journal} {Phys. Rev. E}\ }\textbf {\bibinfo {volume} {95}},\ \bibinfo
  {pages} {052135} (\bibinfo {year} {2017})}\BibitemShut {NoStop}%
\bibitem [{\citenamefont {Swendsen}()}]{Swendsen_2017}%
  \BibitemOpen
  \bibfield  {author} {\bibinfo {author} {\bibfnamefont {R.~H.}\ \bibnamefont
  {Swendsen}},\ }\href@noop {} {\bibinfo  {journal} {arXiv:1702.05810}\
  }\BibitemShut {NoStop}%
\bibitem [{\citenamefont {Franzosi}(2011)}]{Franzosi_JSP11}%
  \BibitemOpen
\bibfield  {journal} {  }\bibfield  {author} {\bibinfo {author} {\bibfnamefont
  {R.}~\bibnamefont {Franzosi}},\ }\href {\doibase 10.1007/s10955-011-0200-4}
  {\bibfield  {journal} {\bibinfo  {journal} {Journal of Statistical Physics}\
  }\textbf {\bibinfo {volume} {143}},\ \bibinfo {pages} {824} (\bibinfo {year}
  {2011})}\BibitemShut {NoStop}%
\bibitem [{\citenamefont {Franzosi}(2012)}]{Franzosi_PRE12}%
  \BibitemOpen
  \bibfield  {author} {\bibinfo {author} {\bibfnamefont {R.}~\bibnamefont
  {Franzosi}},\ }\href {\doibase 10.1103/PhysRevE.85.050101} {\bibfield
  {journal} {\bibinfo  {journal} {Phys. Rev. E}\ }\textbf {\bibinfo {volume}
  {85}},\ \bibinfo {pages} {050101(R)} (\bibinfo {year} {2012})}\BibitemShut
  {NoStop}%
\bibitem [{\citenamefont {Rugh}(1997)}]{RughPRL97}%
  \BibitemOpen
  \bibfield  {author} {\bibinfo {author} {\bibfnamefont {H.~H.}\ \bibnamefont
  {Rugh}},\ }\href {\doibase 10.1103/PhysRevLett.78.772} {\bibfield  {journal}
  {\bibinfo  {journal} {Phys. Rev. Lett.}\ }\textbf {\bibinfo {volume} {78}},\
  \bibinfo {pages} {772} (\bibinfo {year} {1997})}\BibitemShut {NoStop}%
\bibitem [{\citenamefont {Federer}(1969)}]{Federer_1969}%
  \BibitemOpen
  \bibfield  {author} {\bibinfo {author} {\bibfnamefont {H.}~\bibnamefont
  {Federer}},\ }\href@noop {} {\emph {\bibinfo {title} {Geometric Measure
  Theory}}}\ (\bibinfo  {publisher} {Die Grundlehren der mathematischen
  Wissenschaften, Band 153, New York: Springer-Verlag New York Inc.},\ \bibinfo
  {year} {1969})\ p.\ \bibinfo {pages} {676}\BibitemShut {NoStop}%
\bibitem [{\citenamefont {Laurence}(1989)}]{Laurence_1989}%
  \BibitemOpen
  \bibfield  {author} {\bibinfo {author} {\bibfnamefont {P.}~\bibnamefont
  {Laurence}},\ }\href {\doibase 10.1007/BF00945002} {\bibfield  {journal}
  {\bibinfo  {journal} {Zeitschrift für angewandte Mathematik und Physik
  ZAMP}\ }\textbf {\bibinfo {volume} {40}},\ \bibinfo {pages} {258} (\bibinfo
  {year} {1989})}\BibitemShut {NoStop}%
\bibitem [{\citenamefont {Franzosi}\ \emph {et~al.}(1999)\citenamefont
  {Franzosi}, \citenamefont {Casetti}, \citenamefont {Spinelli},\ and\
  \citenamefont {Pettini}}]{Franzosi_PRE99}%
  \BibitemOpen
  \bibfield  {author} {\bibinfo {author} {\bibfnamefont {R.}~\bibnamefont
  {Franzosi}}, \bibinfo {author} {\bibfnamefont {L.}~\bibnamefont {Casetti}},
  \bibinfo {author} {\bibfnamefont {L.}~\bibnamefont {Spinelli}}, \ and\
  \bibinfo {author} {\bibfnamefont {M.}~\bibnamefont {Pettini}},\ }\href
  {\doibase 10.1103/PhysRevE.60.R5009} {\bibfield  {journal} {\bibinfo
  {journal} {Phys. Rev. E}\ }\textbf {\bibinfo {volume} {60}},\ \bibinfo
  {pages} {R5009} (\bibinfo {year} {1999})}\BibitemShut {NoStop}%
\bibitem [{\citenamefont {Franzosi}\ \emph {et~al.}(2000)\citenamefont
  {Franzosi}, \citenamefont {Pettini},\ and\ \citenamefont
  {Spinelli}}]{Franzosi_PRL00}%
  \BibitemOpen
  \bibfield  {author} {\bibinfo {author} {\bibfnamefont {R.}~\bibnamefont
  {Franzosi}}, \bibinfo {author} {\bibfnamefont {M.}~\bibnamefont {Pettini}}, \
  and\ \bibinfo {author} {\bibfnamefont {L.}~\bibnamefont {Spinelli}},\ }\href
  {\doibase 10.1103/PhysRevLett.84.2774} {\bibfield  {journal} {\bibinfo
  {journal} {Phys. Rev. Lett.}\ }\textbf {\bibinfo {volume} {84}},\ \bibinfo
  {pages} {2774} (\bibinfo {year} {2000})}\BibitemShut {NoStop}%
\bibitem [{\citenamefont {Buonsante}\ \emph {et~al.}(2004)\citenamefont
  {Buonsante}, \citenamefont {Penna},\ and\ \citenamefont
  {Vezzani}}]{BPV_PRB04}%
  \BibitemOpen
  \bibfield  {author} {\bibinfo {author} {\bibfnamefont {P.}~\bibnamefont
  {Buonsante}}, \bibinfo {author} {\bibfnamefont {V.}~\bibnamefont {Penna}}, \
  and\ \bibinfo {author} {\bibfnamefont {A.}~\bibnamefont {Vezzani}},\ }\href
  {\doibase 10.1103/PhysRevB.70.184520} {\bibfield  {journal} {\bibinfo
  {journal} {Phys. Rev. B}\ }\textbf {\bibinfo {volume} {70}},\ \bibinfo
  {pages} {184520} (\bibinfo {year} {2004})}\BibitemShut {NoStop}%
\bibitem [{\citenamefont {Franzosi}\ \emph {et~al.}(2010)\citenamefont
  {Franzosi}, \citenamefont {Giampaolo},\ and\ \citenamefont
  {Illuminati}}]{Franzosi_PRA10}%
  \BibitemOpen
  \bibfield  {author} {\bibinfo {author} {\bibfnamefont {R.}~\bibnamefont
  {Franzosi}}, \bibinfo {author} {\bibfnamefont {S.~M.}\ \bibnamefont
  {Giampaolo}}, \ and\ \bibinfo {author} {\bibfnamefont {F.}~\bibnamefont
  {Illuminati}},\ }\href {\doibase 10.1103/PhysRevA.82.063620} {\bibfield
  {journal} {\bibinfo  {journal} {Phys. Rev. A}\ }\textbf {\bibinfo {volume}
  {82}},\ \bibinfo {pages} {063620} (\bibinfo {year} {2010})}\BibitemShut
  {NoStop}%
\end{thebibliography}%

\end{document}